\documentclass{article}

\def\be{\begin{equation}}
\def\ee{\end{equation}}
\def\bea{\begin{eqnarray}}
\def\eea{\end{eqnarray}}

\usepackage{caption}
\usepackage{lineno}
\usepackage{graphicx}
\usepackage{placeins}
\usepackage{float}
\usepackage{subfigure}

\title{The MURAVES experiment: study of the Vesuvius Great Cone with Muon Radiography}
\author{M. D'Errico}
\author{M. D'Errico$^{1,2}$ \and F. Ambrosino $^{1,2}$ \and A. Anastasio$^2$ \and S. Basnet $^3$ \and L. Bonechi $^4$\and M. Bongi$^{4,5}$\and  A. Bross $^{6}$\and R. Ciaranfi$^4$\and  L. Cimmino$^{1,2}$\and  C. Ciulli$^{4,5}$\and R. D'Alessandro$^{4,5}$\and A. Giammanco$^3$\and F. Giudicepietro$^7$\and S. Gonzi$^{4,5}$\and R. Karnam$^3$ \and G. Macedonio$^7$\and V. Masone$^2$\and N. Mori$^{4,5}$\and M. Moussawi$^3$\and M. Orazi$^7$\and G. Passeggio$^2$ \and R. Peluso$^7$\and A. Pla- Dalmau$^6$\and C. Rendon$^8$\and A. Samalan$^8$\and G. Saracino$^{1,2}$\and G. Scarpato$^7$\and P. Strolin$^{1,2}$\and M. Tytgat$^8$\and E. Vertechi$^7$ \and L.Viliani$^{4,5}$. }
\date{ %
$^1$ University of Naples Federico II, Naples, Italy \\ 
$^2$ INFN sez. di Napoli, Naples, Italy \\
$^3$ Centre for Cosmology, Particle Physics and Phenomenology, Université catholique de Louvain, Louvain-la-Neuve, Belgium \\
$^4$ INFN sez. di Firenze, Florence, Italy \\
$^5$ University of Florence, Florence, Italy \\
$^6$ Fermilab, Batavia, IL, USA \\
$^7$INGV, Osservatorio Vesuviano, Naples, Italy \\ 
$^8$ Department of Physics and Astronomy, Ghent University, Ghent, Belgium\\
}

\begin{document}
\maketitle

\begin{abstract}
The MURAVES experiment aims at the muographic imaging of
the internal structure of the summit of Mt. Vesuvius,
exploiting muons produced by cosmic rays.
Though presently quiescent, the volcano carries a dramatic
hazard in its highly populated surroundings. The
challenging measurement of the rock density distribution
in its summit by muography, in conjunction with data from
other geophysical techniques, can help the modeling of
possible eruptive dynamics.
The MURAVES apparatus consists of an array of three
independent and identical muon trackers, with a total
sensitive area of 3 square meters. In each tracker, a sequence of 4
XY tracking planes made of plastic scintillators is
complemented by a 60 cm thick lead wall inserted between
the two downstream planes to improve rejection of
background from low energy muons. The apparatus is
currently acquiring data. Preliminary results from the
analysis of a first data sample are presented.
\end{abstract}

\section{Introduction}
Mt. Vesuvius is an explosive volcano, still active though quiescent since about 80 years and located near the city of Naples. Possible eruptions would be particularly dangerous due to the dense urbanization even in its close surroundings. The history of Vesuvius is characterized by several violent eruptions. The best known is the Plinian eruption of 79 b. C. that destroyed the city of Pompeii and other settlements around it.  \\ 
In the course of millennia the morphology of the volcano strongly changed due to its activity. In particular, several explosive eruptions in the last period of activity caused the collapse of the caldera and the formation of the present crater. The interior of the cone is thought to be composed of materials having different densities, probably distributed in a layered structure. \\
So far muon radiography (or "muography") was applied to volcanoes requiring muon penetration through a rock thickness below 1 km. However, exploring the mass distribution inside the summit of Vesuvius' cone roughly requires doubling the extent of the muon range, which implies coping with very low muon rates and unexperienced backgrounds. This is the  challenge of the MURAVES (MUon RAdiography of VESuvius) project that is described in the following, concluding with the presentation of some preliminary results.\\ 

\section{Technology and Infrastructure}

\begin{figure}
\centering
\includegraphics[width=6.2cm]{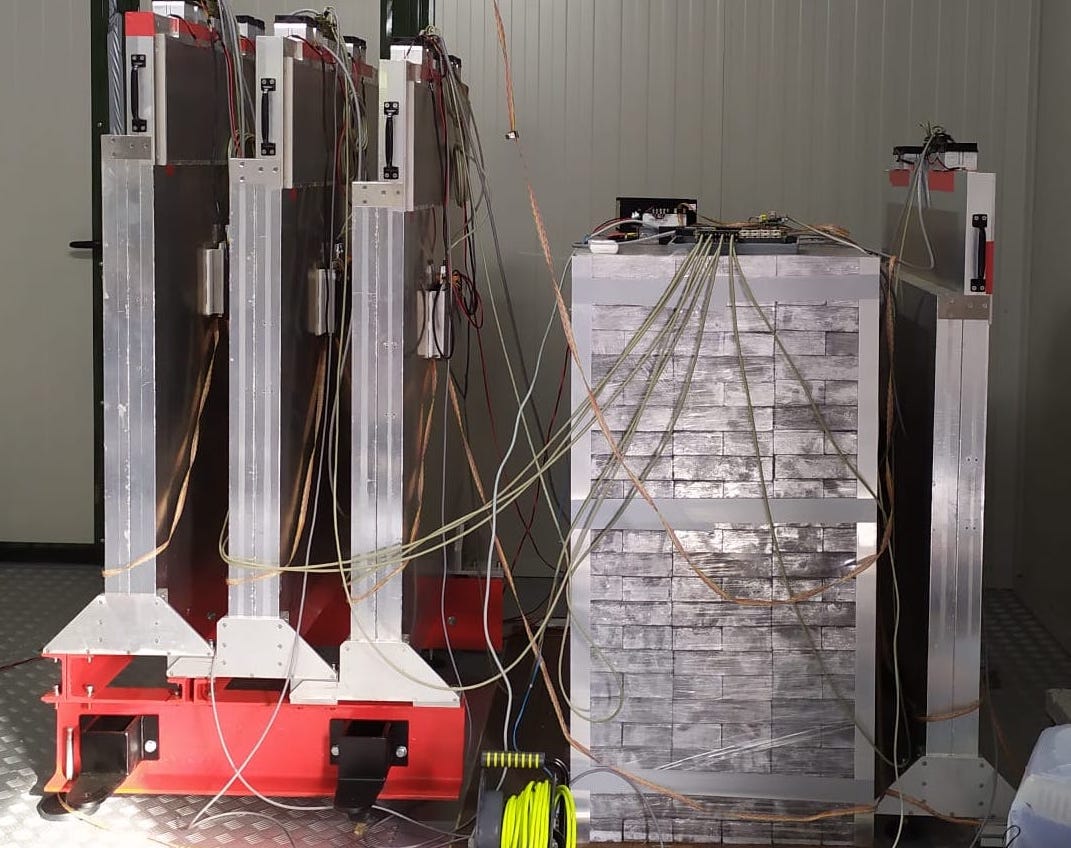}
\includegraphics[width=5cm]{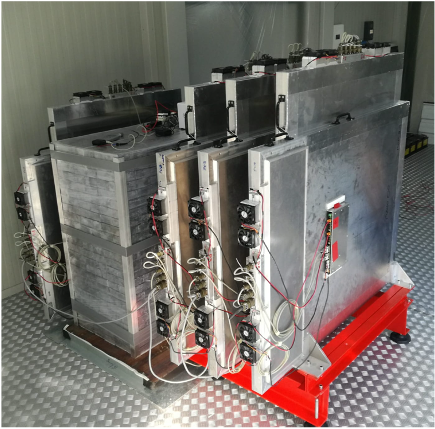}
\caption{One of the three independent muon trackers.}
\label{fig:Hodoscope}
\end{figure}

\begin{figure}[!htp]
\centering
\includegraphics[width=11.1cm]{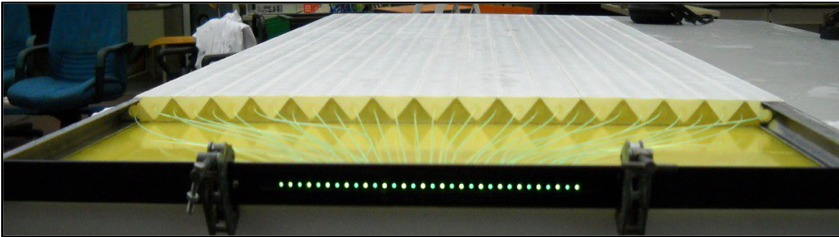}
\includegraphics[width=5.7cm]{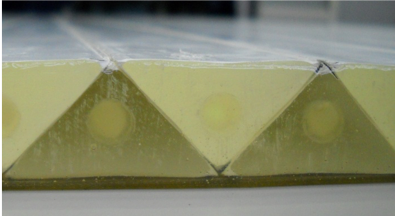}
\caption{Top: The 32 scintillator bars and WLS fibers assembled in a half of a planar array. Bottom: triangular section of the scintillator bars.}
\label{fig:bars}
\end{figure}

\begin{figure}
\centering
\includegraphics[width=6 cm]{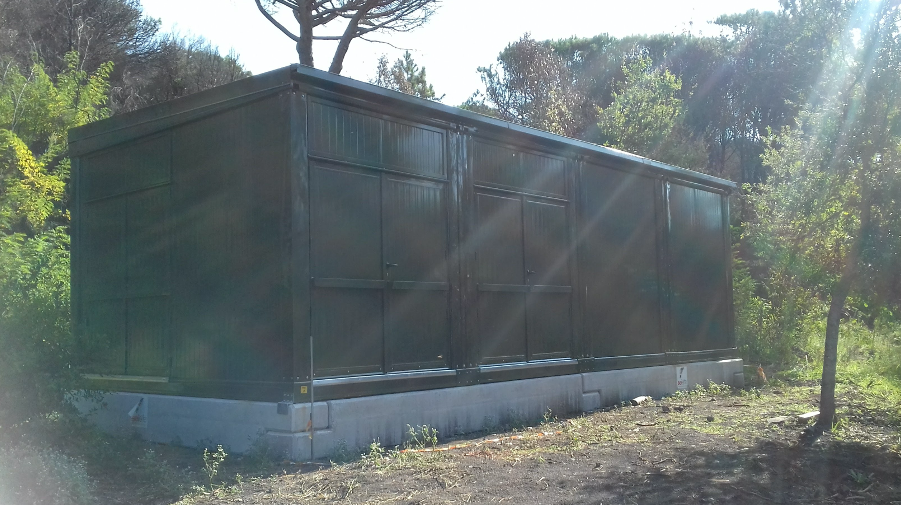}
\includegraphics[width=6 cm]{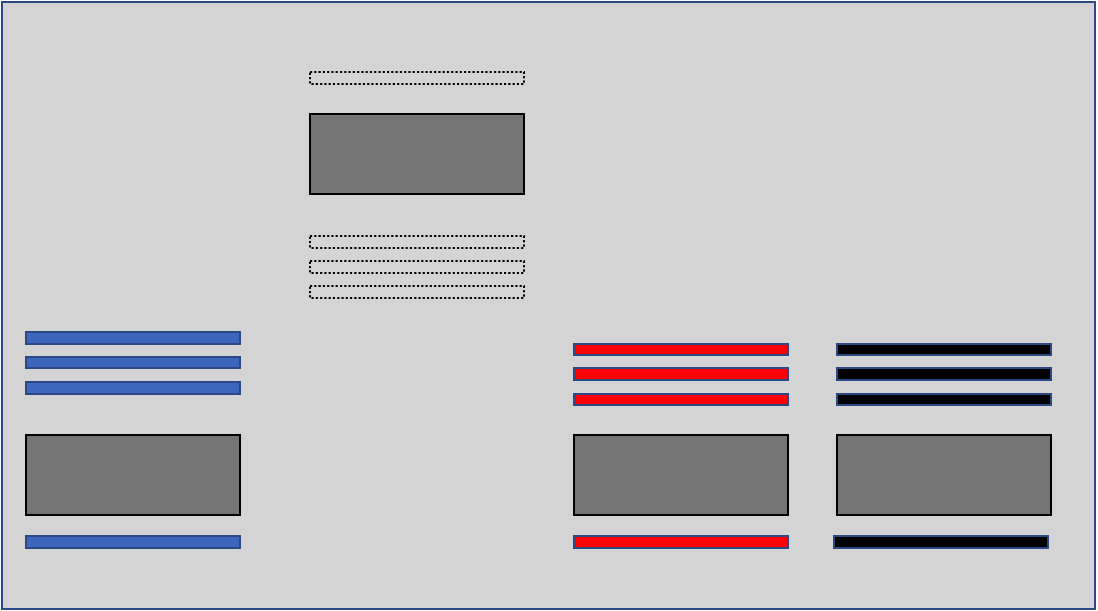}
\caption{Left: The container installed at Mt. Vesuvius. Right: Scheme of the arrangement of the muon trackers inside the container: the not-colored tracker indicates the position of a muon tracker while taking calibration data.}
\label{fig:container}
\end{figure}

The MURAVES muon telescope has a total sensitive cross-sectional area of 3 m$^2$, provided by an array of three identical muon trackers. Each muon tracker is formed by a sequence of four tracking stations distributed over a length of 2 meters, with a 60 cm thick lead wall inserted between the two downstream tracking stations for the purpose of background reduction, in particular tagging and rejecting low energy muons. The tracking stations, realized on the basis of the \textit{ Mu-Ray} technology \cite{Ambrosi_MuRay} \cite{Anastasio_MuRay} \cite{Ambrosino_MuRay}, consist of two adjacent planar arrays of plastic scintillator bars, orthogonally oriented to provide the horizontal and vertical coordinates of the muon impact point. Figure \ref{fig:Hodoscope} shows one of the three muon trackers. The bars have a cross-section shaped as an isosceles triangle with 1.7 cm height and 3.3 cm basis (Figure \ref{fig:bars}) and are assembled with alternate up-down orientation of the triangle. The space resolution of the corresponding coordinate is improved through the weighted average of the signal in adjacent bars. The plastic scintillator bars were extruded at the FERMILAB-NICADD facility from a bulk of polystyrene with the addition of PPO and POPOP scintillating dopants, emitting in a blue wavelength spectrum centered on a 420 nm wavelength. They are of the same type as those used in the D0 \cite{D0} and Minerva  \cite{Minerva} experiments. Multi-clad Kuraray Y11 S-35 wavelength shifting (WLS) fibers with a 1.2 mm diameter run in a co-extruded central hole of 1.5 $\pm$ 0.1 mm diameter. Their absorption and emission spectra are in the blue wavelength range 400-470 nm and in the green range 470-550 nm, respectively. The WLS fibers are read-out by ASD-RGB1C-P Silicon Photo-Multipliers (SiPM), produced by the Advansid company \cite{Advansid}.\\
\begin{figure}[!htp]
\centering
\includegraphics[width=5.45 cm]{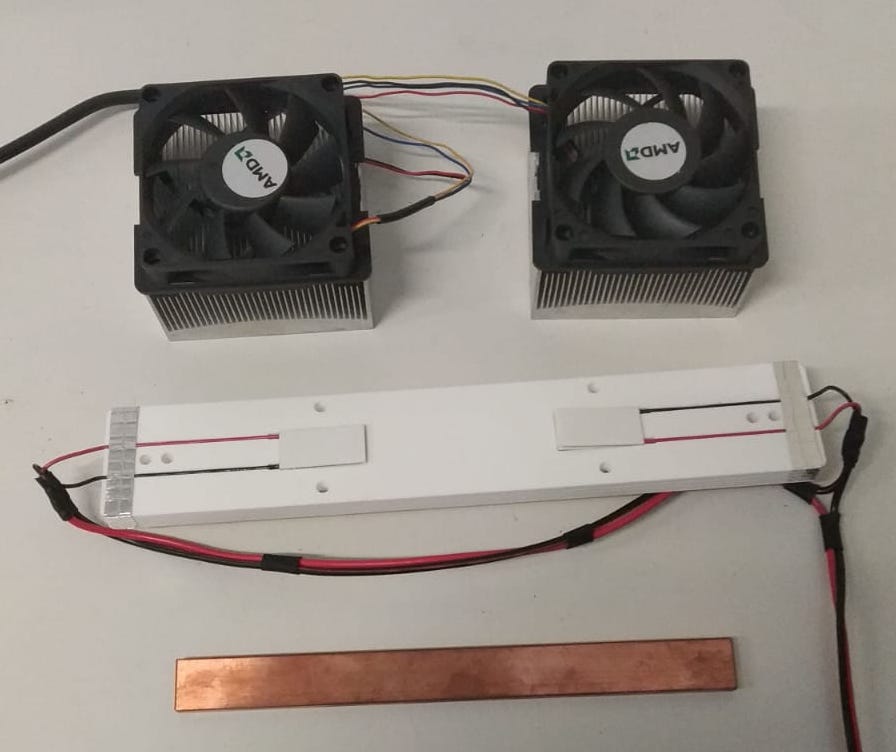}
\includegraphics[width= 5 cm]{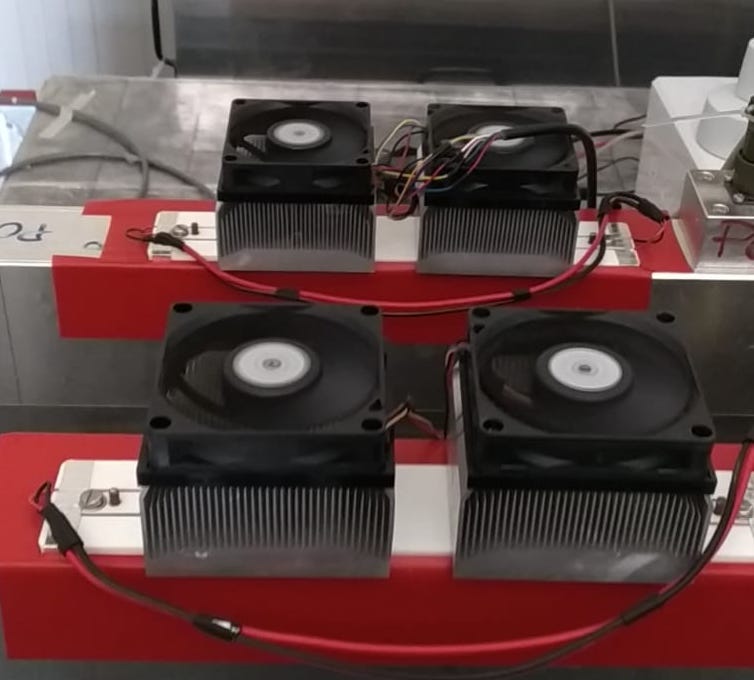}
\caption{The temperature control system: the components (left); the system installed on the detector (right).}
\label{fig:TempSystem}
\end{figure}
Each muon tracker has 8 planar arrays of plastic scintillator bars. The 32 SiPMs of each half array (16 half-arrays for each muon tracker) are hosted on a hybrid circuit and are connected to a 32-channel front-end electronics (FEE) based on the ASIC chip EASIROC \cite{EASIROC_ref}. The setting of the trigger and the data acquisition of each muon tracker is provided by 16 slave boards managed by a master board. The readout and data acquisition system is described in \cite{Cimmino_2017}. The experimental apparatus and the data transfer are remotely controlled. 
\\As they are solid state devices, the SiPMs operation is affected by temperature variations. A custom designed temperature control system based on Peltier cells cools or heats the photosensors according to the requirement. For each group of 32 SiPMs it consists of two Peltier cells, a copper strip providing an uniform thermal conduction and a couple of fans for heat dissipation. A custom circuit controls the cells through Pulse Width Modulation drivers. The temperature can be set within 5-7 $^\circ$C maximum difference with respect to the ambient temperature. The bias voltage and the temperature determine the working point of the SiPMs. The stability of the SiPMs performance is thus obtained by changing the bias voltage according to the temperature, so as to maintain a constant overvoltage. The temperature control system is shown in Figure \ref{fig:TempSystem}. Anyhow, the SiPMs must be operated at a temperatures resulting in a sufficiently low power consumption and far enough from the dew point, to prevent damage from condensation.\\
The MURAVES telescope is housed inside a container, installed on the south-west  slope  of Mt. Vesuvius at 600 m a.s.l. and 1500 m distance from the summit. The electric power is supplied by a solar panel system located on the roof of the container and connected to an array of batteries, which ensures continuity at night or cloudy days. Four concrete platforms inside the container rest directly on the ground and can support lead walls of up to 90 cm thickness (at present 60 cm). One of the platforms is intended to take calibration data with trackers in turn pointing to free sky opposite to the mountain. Figure \ref{fig:container} shows the container and a scheme of the muon trackers' arrangement inside the container. 

\section{Measurement and sensitivity} 
\begin{figure}
\centering
\includegraphics[width=12cm]{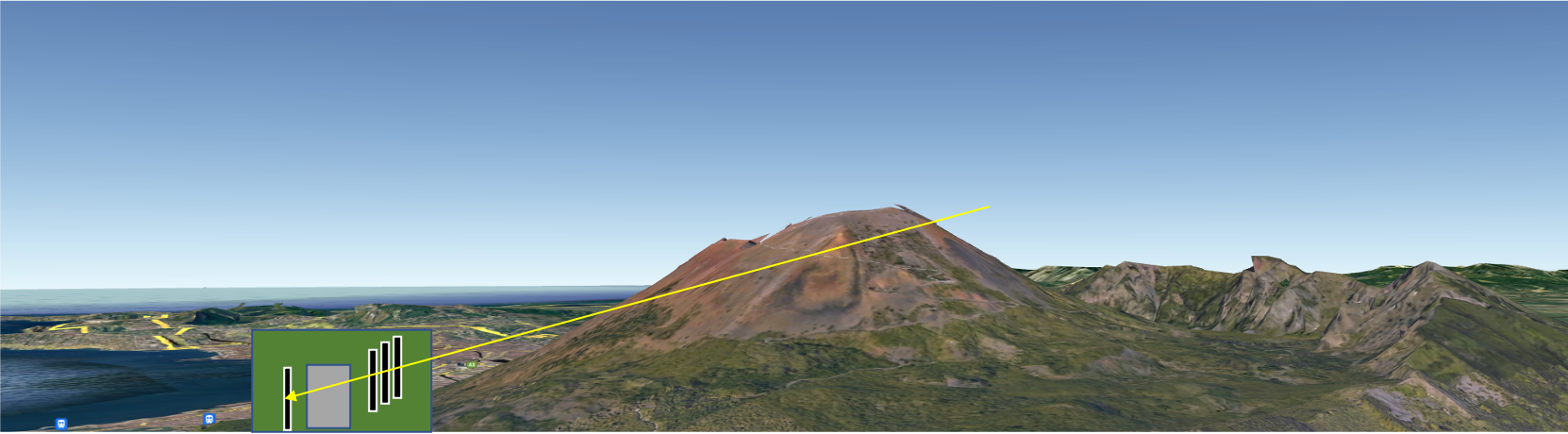}
\includegraphics[width=12cm]{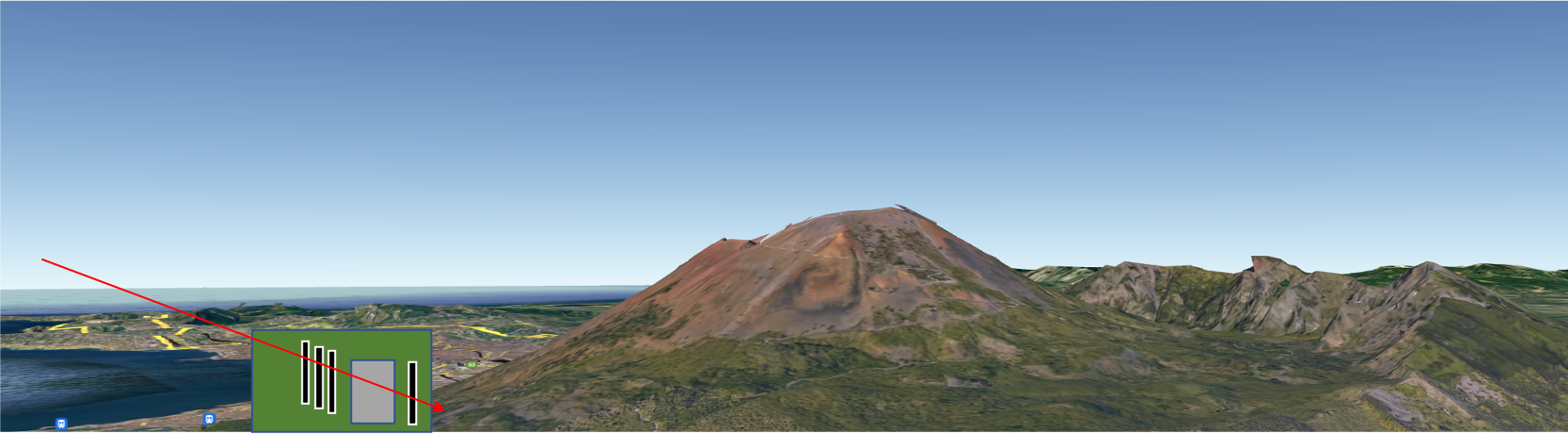}
\caption{Scheme of the configuration for the measurement of the muon flux transmitted through the volcano (top figure) and of  the free-sky calibration flux (bottom figure). Detector not in scale. }
\label{fig:VesuvioMode}
\end{figure}

\begin{figure} 
\centering
\includegraphics[width=12 cm]{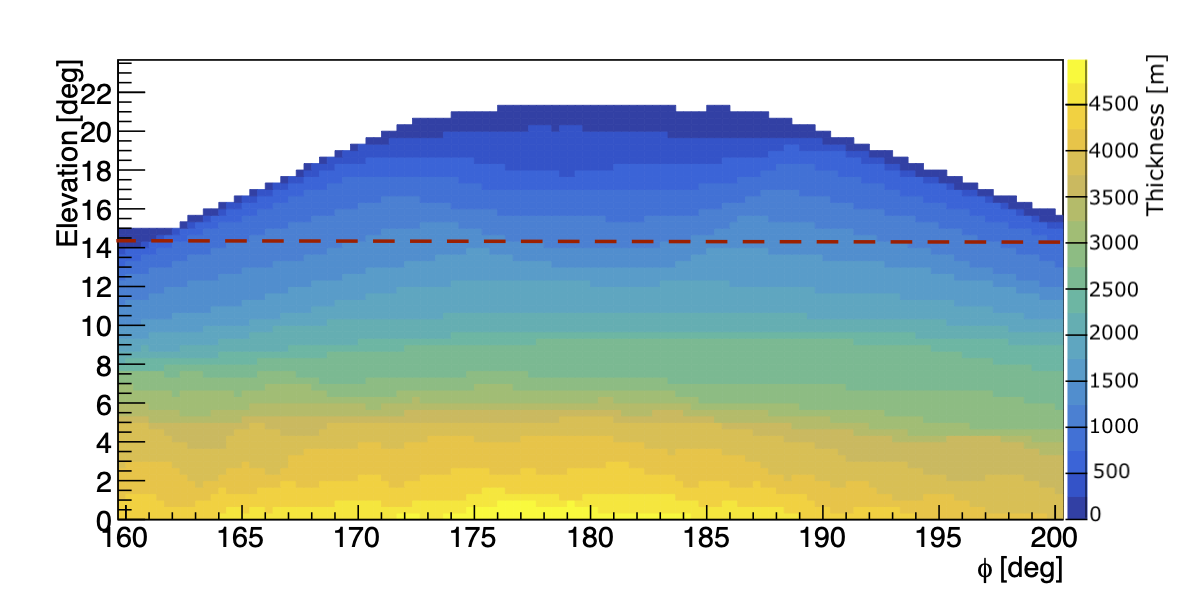}
\caption{Map of the rock thickness to be traversed by muons through Mt. Vesuvius as a function of the elevation and horizontal angles. The horizontal range reflects the detector's geometrical acceptance. }
\label{Vesuvius_thickness} 
\end{figure}

The experiment aims at mapping the mean density of the matter crossed by muons in the traversal of the volcano, through the measurement of the muon flux that reaches the detector. Its ratio with the muon flux measured in calibration runs with the muon tracker pointing to open sky in the opposite direction (figure \ref{fig:VesuvioMode}) gives the \textit{muon transmission}:
\begin{equation}
T (\theta,\phi) = \frac{N_{\mu}^{v} (\theta, \phi) / \Delta t^v} {N_{\mu}^{fs} (\theta, \phi)/ \Delta t^{fs}} = \frac{ \epsilon^v \cdot S_{eff}(\theta,\phi) \int_{E_{min(\rho)}}^\infty \Phi(\theta, \phi; E) dE }{\epsilon^{fs} \cdot S_{eff}(\theta,\phi) \int_{E_0}^\infty \Phi (\theta,\phi;E) dE} 
\end{equation}
where $\Phi (\theta, \phi; E) $ is the  differential muon flux,  $N_{\mu}^{v} (\theta, \phi)$ and $N_{\mu}^{fs} (\theta, \phi$ are the muons observed in a time interval $\Delta t^v$ and $\Delta t^{fs}$ respectively,  with the apices $v$ and $fs$ indicating \textit{volcano} and \textit{free-sky} data. $E_{min(\rho)}$ is the energy needed to survive the rock and to be seen by the detector; $E_0$ is the minimum energy needed to not be absorbed in the detector itself. The measured fluxes can be factorized as indicated on the r.h.s. so that with a sufficiently good approximation the efficiency  $\epsilon$ and the effective sensitive area  $S_{eff}$ of the muon tracker cancel out, isolating the ratio of the muon fluxes $\Phi (\theta, \phi; E) $ as a function of the elevation and azimuthal angles $\theta$ and $\phi$.\\
Figure \ref{Vesuvius_thickness} shows the rock thickness to be traversed by muons, as evaluated exploiting a Digital Elevation Model (DEM) of Mt. Vesuvius \cite{DEM_ref}. The rock thickness ranges from some hundreds meters at the uppermost part of the cone to almost 5000 meters at its basis. The sensitivity of the experiment depends on the statistics which can be accumulated and on the background which is left in spite of the tools for its reduction.
\begin{table}[!htbp]
\centering
\begin{tabular}{l | l | l| l| l}
$\Delta$y & $\Delta$x & $\bar{L}$ = 500 m & $\bar{L}$ = 1000 m & $\bar{L}$ = 3000 m \\
\hline
9 m & 9 m & 8 months & 3 years & 100 years\\
9 m & 26 m & 3 months & 1 year & 33 years \\
9 m & 130 m & 15 days & 2.5 months & 6 years \\
26 m & 130 m & 5 days &  1 month & 2 years \\
52 m & 260 m & 2 days & 6 days & 16 months \\
\hline
\end{tabular} 
\caption{Exposure times  expected to be required to measure the mean density with a 10 \% statistical uncertainty for a set of values of the muon path length in the rock and of horizontal and vertical space resolutions $\Delta $x and $\Delta$y.}
\label{tab:Expected_exposuretimes}
\end{table}
The expected muon flux through the volcano was evaluated using simulations. Table \ref{tab:Expected_exposuretimes} gives estimates of the exposure times required to measure the mean rock density $\rho$ with a statistical uncertainty $\Delta\rho / \rho = 10\%$, for a set of volumes defined by $\Delta x $ and $\Delta y$ and the rock thicknesses $L$, averaged to $\bar{L}$. The required exposure times have been obtained as described in \cite{Lesparre_2010}. The nominal density has been taken to be  $\rho$ =  2.65 g cm$^{-3}$. In a first approximation, data are assumed to be background free and, presumably of a lesser importance, the detector efficiency were not  taken into account.  Looking at  the thickness map of Mt. Vesuvius in Figure \ref{Vesuvius_thickness} in the light of  Table \ref{tab:Expected_exposuretimes} one can deduce that a detector of a few m$^2$ area like MURAVES can face the challenge of exploring the summit of the cone of Mt. Vesuvius. A larger area is required for a path length in the 3 km range. 

\section{Datasets and data analysis} 
\begin{figure}
\centering
\includegraphics[width=8 cm]{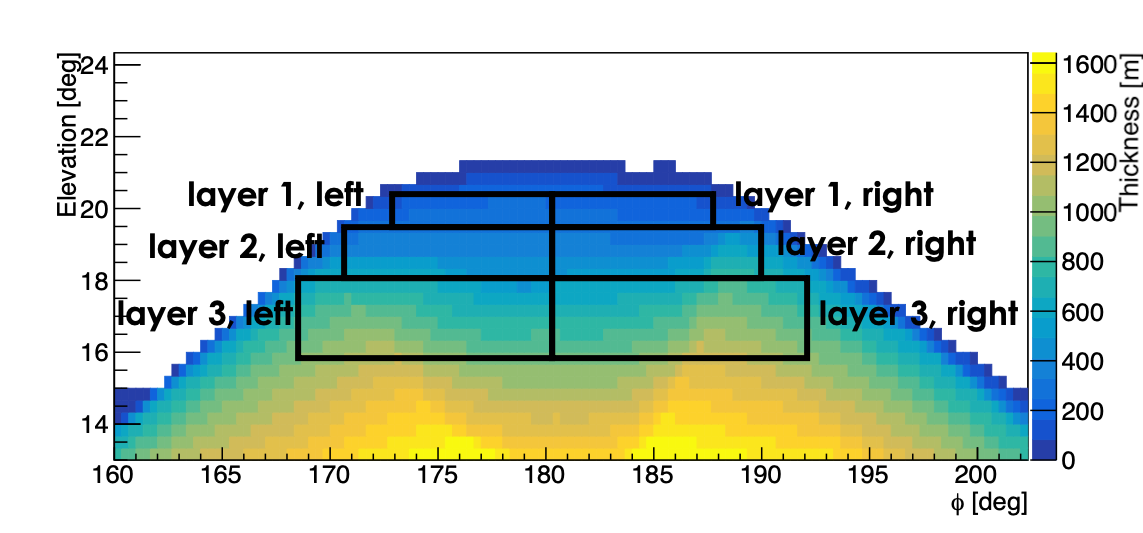}
\includegraphics[width=4 cm]{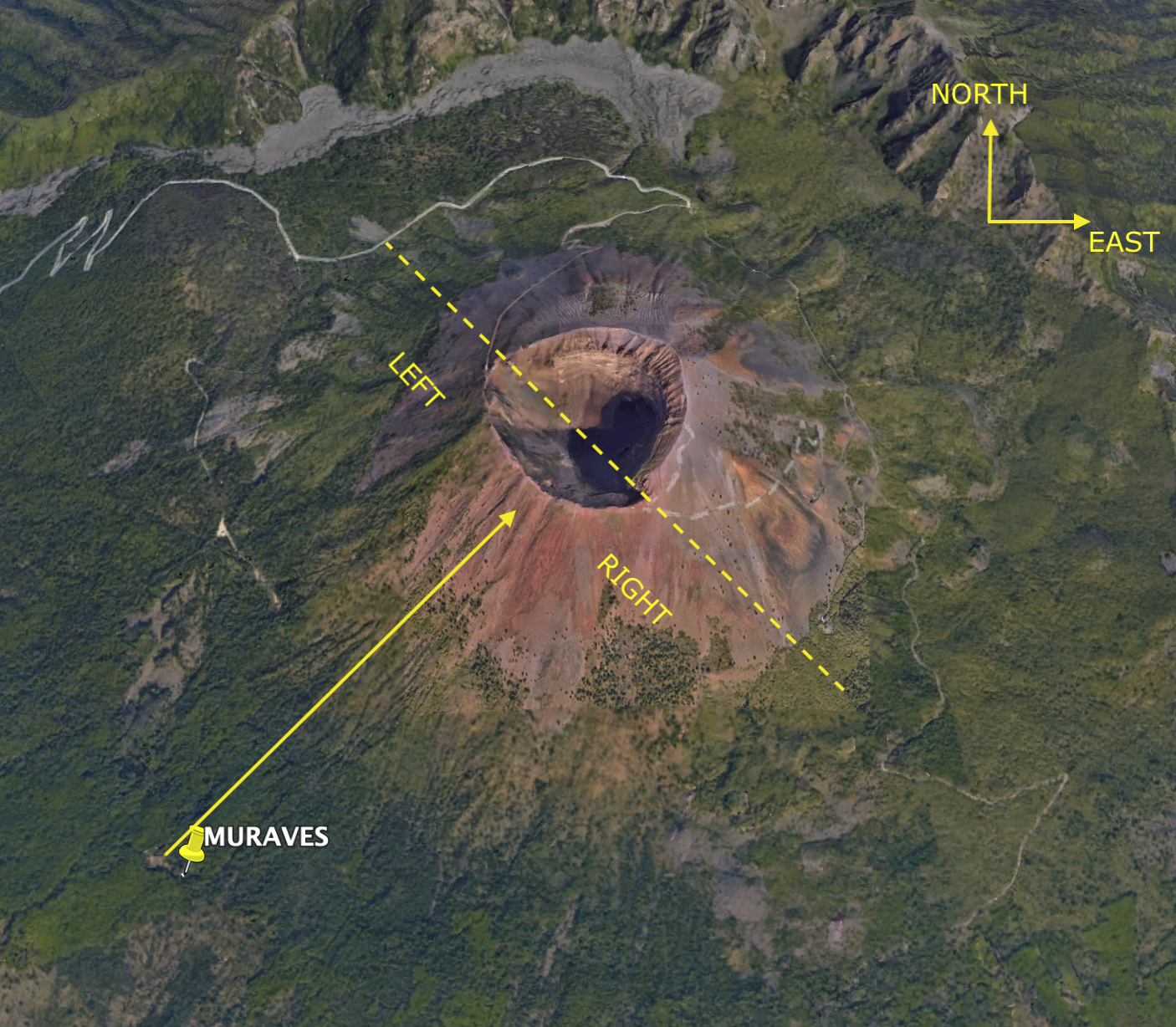}
\caption{Left: Map of the muon path length through the volcano, with indication of the angular regions where the average density is at present in the process to be evaluated. Right: map of the Vesuvius area with the indication of the detector orientation. }
\label{Highilight_regions}
\end{figure}

\begin{figure}
\centering
\includegraphics[width= 12 cm]{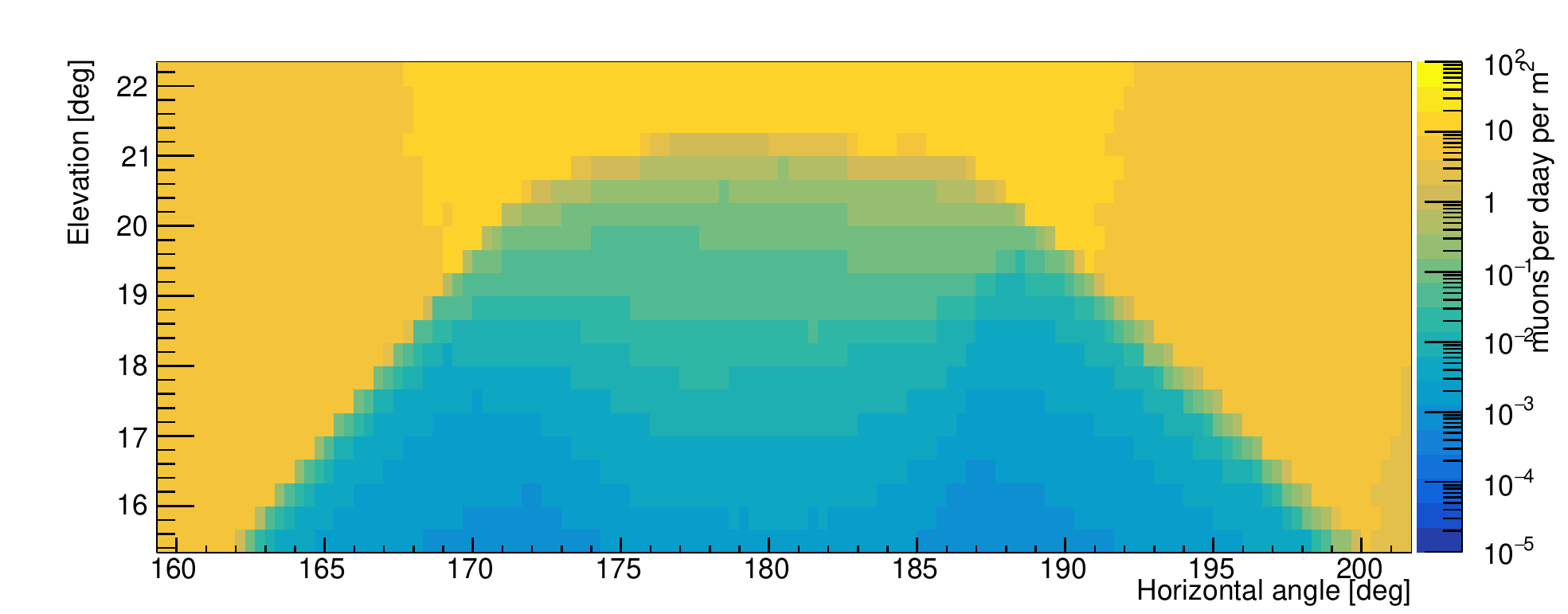}
\includegraphics[width= 12 cm]{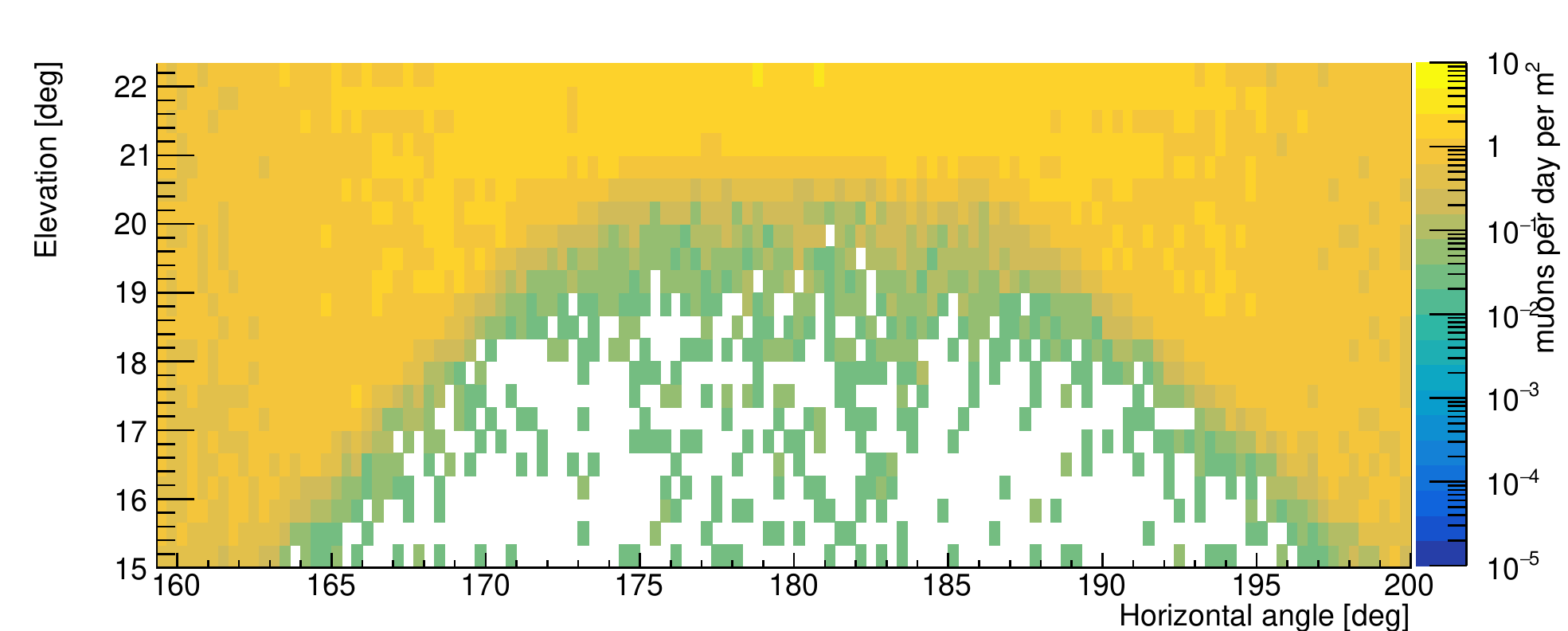}
\caption{Expected Flux, evaluated with PUMAS (left); Measured Flux obtained from one of the datasets in Table \ref{VesuviusAcquistionTimes}. }
\label{flux_comparison}
\end{figure}

Data adequate for a preliminary analysis was acquired by two of the three muon trackers of the MURAVES telescope, named NERO and ROSSO. Because of the variations in the environmental conditions, the SiPMs were operated at two different  working points corresponding to temperatures 15$^\circ$C and 20$^\circ$C. Correspondingly, data is subdivided in the four groups shown  in Table \ref{VesuviusAcquistionTimes}. The table  gives the respective  effective exposures. The exposures pointing to the volcano quoted in the table have an order of magnitude of 1-2 months, to be compared to a few years of data taking foreseen for MURAVES, with the full muon telescope. The table gives also the effective exposure already available for each free-sky calibration of the datasets. The free-sky exposures are shorter, but thanks to the high muon flux they give a comparable or even lower statistical uncertainty. \\
\begin{table}[!htbp]
\centering
\begin{tabular}{c|c|c}
Dataset & Vesuvius & Free-sky  \\
\hline
ROSSO wp 15$^\circ$C  & 51 days & 9.5 days\\
ROSSO wp 20$^\circ$C &  40 days & 14.3 days\\
NERO wp 15$^\circ$C & 43 days  & 10 days\\
NERO wp 20$^\circ$C & 26 days & 17 days\\
\hline
\end{tabular}
\caption{Exposure times of the four datasets with the two muon trackers pointing to Mt. Vesuvius or to free-sky for calibration.}
\label{VesuviusAcquistionTimes}
\end{table}
A preliminary analysis was performed separately for the four datasets. After a selection of the good muon events, the data analysis proceeded by dividing the angular range corresponding to the upper part of the cone into regions large enough to contain a reasonable statistics, as shown in Figure \ref{Highilight_regions}. These regions were chosen to be  larger in azimuthal angle than in elevation, so as to obtain a better visibility of a possible stratification of materials of different densities. For a better balance of the rates, the layers cover an increasing range of elevation angles going down from the top of the volcano. The three layers were further subdivided in the left and right parts as shown in the figure (west and a east side, respectively) in order to show possible asymmetries in principle visible from the detector location.
The number of events registered in each of the six chosen angular regions are listed in Table \ref{statistics} separately for each of the four datasets. The statistical uncertainty is maintained below 10\% in most cases, in a few cases is larger but not exceeding 20\%.
\begin{table}[!htp]
\centering
\begin{tabular}{l|c|c|c|c}
 &  \multicolumn{2}{c}{N events} &\multicolumn{2}{c}{Stat. unc. (\%)} \\
\hline
 Dataset & left & right & left & right \\ 
  \hline
   \multicolumn{5}{c}{Layer 1} \\  
   \hline
 ROSSO wp 15 $^\circ$C & 428 & 439 & 0.05 & 0.05 \\
 ROSSO wp 20 $^\circ$C & 346 & 323 & 0.05 & 0.05 \\
 NERO wp 15 $^\circ$C & 231 & 258 & 0.05 & 0.05 \\
 NERO wp 20 $^\circ$C & 128 & 258 & 0.07 & 0.06 \\
 \hline
  \multicolumn{5}{c}{Layer 2} \\  
   \hline
 ROSSO wp 15 $^\circ$C & 164 & 140 & 0.08 & 0.08 \\
 ROSSO wp 20 $^\circ$C & 106  & 109 & 0.10 & 0.10 \\
 NERO wp 15 $^\circ$C & 78 & 79 & 0.11 & 0.11 \\
 NERO wp 20 $^\circ$C & 61 & 63 & 0.13 & 0.13 \\
 \hline
\multicolumn{5}{c}{Layer 3} \\  
  \hline
 ROSSO wp 15 $^\circ$C & 61 & 76 & 0.12 & 0.11 \\
 ROSSO wp 20 $^\circ$C & 58  & 63 & 0.13 & 0.13 \\
 NERO wp 15 $^\circ$C & 47 & 47 & 0.14 & 0.15 \\
 NERO wp 20 $^\circ$C & 27 & 30 & 0.19 & 0.18 \\
 \hline
\end{tabular}
\caption{Events and statistical uncertainties in the six angular regions, separately for the four datasets.}
\label{statistics}
\end{table} 
The expected muon flux through the volcano was evaluated using the PUMAS software  \cite{PUMAS}, a backward Montecarlo simulation program especially thought for muography applications. The muon absorption in the volcano and the lead wall energy threshold were taken into account. Figure \ref{flux_comparison} compares the expected muon flux to the measured flux given by one of the datasets, to be taken as an example. Within uncertainties there is agreement. The simulation chain is still under development and a higher level of details is being included. In particular, the detector responses and the effects of analysis cuts for background reduction, such as those on the scattering in the detector and on the muon time-of-flight for forward/backward discrimination. 
The expected transmission will be evaluated with a dedicated software chain, that is currently under construction. Finally the density will be measured in each angular region by comparing the measured transmission to  the expected transmission.
\section{Conclusions}
The MURAVES experiment exploits muography to  investigate, in combination with other techniques, the rock density distribution inside the summit of Mt. Vesuvius, a very dangerous active volcano in the Southern Italy. A muon telescope of 3 m$^2$ total sensitive area, subdivided into three identical and independent muon trackers, were installed on the south-west slope of the volcano inside a container. Four initial datasets are being analyzed, each corresponding to exposures ranging from 26 to 51 days by single muon trackers. Within uncertainties the measured muon agrees with the muon flux expected from a simulation that accounts for the muon absorption in the path through the volcano and for the energy threshold set by the detector. Calibration datasets, taken in the same conditions, exist for each of the four datasets. A first measurement of the mean density of the rock in six angular regions projected  on the volcano is under way. Data taking is foreseen to continue for a few years.
\bibliographystyle{unsrt}

\end{document}